\documentclass[11pt]{article}

\title{Lorentz Invariant Phenomenological Model of Quantum Gravity: A Minimalistic Presentation}

\author{Yuri Bonder}

\date{Instituto de Ciencias Nucleares\\ Universidad Nacional Aut\'onoma de M\'exico\\
A. Postal 70-543, M\'exico D.F. 04510, M\'exico\\ yuri.bonder@nucleares.unam.mx}

\begin{document}
\maketitle

\section*{Abstract}
The purpose of this paper is to give a minimalistic and self-contained presentation of a Lorentz Invariant phenomenological model of Quantum Gravity.

Spacetime, at microscopic scales, may have a non-trivial discrete structure. The goal of Quantum Gravity Phenomenology is to search for empirical manifestations of this structure. In the last years, Quantum Gravity Phenomenology has been dominated by the search of Lorentz Invariance violations (LIV). This is essentially because a discrete spacetime structure may select preferential spacetime directions through an incompatibility with the Lorentz length contraction. However, there are very restricting bounds on LIV \cite{Bounds}. Furthermore, in a certain theory of quantum fields on a discrete spacetime such that the discrete structure determines a preferential direction, it has been shown \cite{Collins} that the radiative corrections would magnify the LIV to levels irreconcilable with observations. These results motivated us to propose a phenomenological model of Quantum Gravity without LIV. In contrast with the phenomenological models with LIV, the starting point is to assume that spacetime building blocks are Lorentz Invariant and look for possible manners in which such building blocks may manifest.

Almost any intuitive discrete structure of spacetime leads to LIV. Therefore, we need to use an analogy\footnote{I thank E. Bianchi for suggesting this analogy.}. Imagine we have a real function $f$ and, for some reason, we need to replace it by a ``discrete'' function made of straight segments of a given length $L$. Then, in the regions where $f$ is straight, the discrete function is indistinguishable from $f$. On the other hand, when $f$ changes dramatically with respect to $L$, the function $f$ is not well approximated with the discrete function. 

We believe that something analogous may occur in Nature: If we assume that spacetime building blocks are Lorentz Invariant, then, according to this analogy, in spacetime regions which are Lorentz Invariant (\textit{i.e.}, flat regions), it is not possible to see the effects of spacetime discreteness. However, in regions where spacetime is curved, its building blocks ``overlap'' and its presence could reveal itself by non-standard interactions of matter and spacetime curvature \cite{QGP}. As the Ricci tensor (and its trace, the curvature scalar) at a point are determined by matter at the same point through Einstein's equations, to couple matter with Ricci tensor may be regarded as a self-coupling, which is not interesting for phenomenological purposes. Thus, we need to search for couplings of matter with Weyl tensor $W_{abcd}$, which can be thought as the Riemann curvature tensor without the Ricci tensor. The coupling terms must be covariant, Lorentz Invariant and with dimensions such that it is not \textit{a priori} suppressed by many powers of a big constant, which is taken to by Planck mass $M_P$.

For practical reasons, we focus in couplings of Weyl tensor and fermionic matters fields, which are generically denoted by $\psi$. Unfortunately, the obvious coupling terms vanish or are highly suppressed by $M_P$, making them phenomenologically uninteresting. For this reason, we use objects having some information of the Weyl tensor but a different index structure, we proceed to construct such objects.

Let $S$ be the space of $2$-forms in spacetime. This space inherits a (pseudo-Riemannian) metric $G$ from spacetime metric $g$, which is given \cite{Wald} by
\begin{equation}
G_{ab cd}=g_{a[c}g_{d]b}.
\end{equation}
Observe that $G_{ab cd}=G_{[ab] [cd]}=G_{cd ab}$ (anti-symmetric pairs of spacetime indexes can be considered as indexes of $S$). The metric $G$ can be used to construct two Hermitian maps from $S$ to $S$ out of the Weyl tensor:
\begin{eqnarray}
{{(W_+)}_{ab}}^{cd}&\equiv&\frac{1}{2}\left({W_{ab}}^{cd}+{W^\dagger_{ab}}^{cd}\right),\\
{{(W_-)}_{ab}}^{cd}&\equiv&\frac{i}{2}\left({W_{ab}}^{cd}-{W^\dagger_{ab}}^{cd}\right),
\end{eqnarray}
where ${W^\dagger_{ab}}^{cd}$ represents the adjoint (with respect to $G_{abcd}$) of ${W_{ab}}^{cd}$. We define $\lambda^{(\pm,l)}$ and $ X_{ab}^{(\pm,l)}=X_{[ab]}^{(\pm,l)}$ such that
\begin{eqnarray}
\label{eigen map} {{(W_\pm)}_{ab}}^{cd} X_{cd}^{(\pm,l)}&=&\lambda^{(\pm,l)} X_{ab}^{(\pm,l)},\\
\label{norm} G_{abcd}X_{ab}^{(\pm,l)}X_{cd}^{(\pm,l)}&=&\pm1,\\
\label{planes} \epsilon_{abcd}X_{ab}^{(\pm,l)}X_{cd}^{(\pm,l)}&=&0.
\end{eqnarray}
Note that the first equation is an eigenvalue equation, $\lambda^{(\pm,l)}$ and  $X_{ab}^{(\pm,l)}$ being, respectively, the eigenvalues and eigenvectors of the Hermitian operators. The label $\pm$ indicates to which operator $\lambda^{(\pm,l)}$ and $X_{ab}^{(\pm,l)}$ are associated and $l=1,2,3$ labels the different eigenvalues (no sum over $l$ is implied if it appears repeatedly in an equation). In $4$ spacetime dimensions, $l$ should run from $1$ to $6$, the number of anti-symmetric pairs of spacetime indexes. However, the identity
\begin{equation}
{\epsilon_{ab}}^{cd} {W_{cd}}^{ef} ={W_{ab}}^{cd} {\epsilon_{cd}}^{ef},
\end{equation}
where $\epsilon_{abcd}$ is spacetime volume $4$-form, implies that there is an unavoidable degeneration in the maps (\ref{eigen map}) which reduces the number of eigenvalues to $3$. This degeneration implies that any linear combination of degenerated eigenvectors is an eigenvector with the same eigenvalue. Conditions (\ref{norm}) and (\ref{planes}) are meant to discriminate from all these linear combinations. This conditions unequivocally fix $X_{ab}^{(\pm,l)}$ unless some eigenvectors are null with respect to $G_{abcd}$. In these case it is not possible to choose the linear combination that the model calls for, therefore, the null eigenvectors are dropped. Moreover, when spacetime is flat, $\lambda^{(\pm,l)}=0$, which is consistent with our motivation. 

Note further that the conditions (\ref{eigen map})-(\ref{planes}) do not determine the sign of  $X_{ab}^{(\pm,l)}$, namely, if $X_{ab}^{(\pm,l)}$ satisfies these conditions, so does $-X_{ab}^{(\pm,l)}$. As the physics is not invariant under $X_{ab}^{(\pm,l)}\rightarrow -X_{ab}^{(\pm,l)}$, we need our coupling to be insensible to that sign. This is achieved by coupling terms which are quadratic in $X_{ab}^{(\pm,l)}$. We then write the most general quadratic term that is minimally suppressed by $M_p$. Let $\widetilde{X}_{ab}^{(\pm,l)}\equiv {\epsilon_{ab}}^{cd}X_{cd}^{(\pm,l)}$, the proposed term has the form
\begin{equation}
L_{int}=H_{ab} \bar{\psi}\gamma^{[a}\gamma^{b]}\psi,
\end{equation}
where
\begin{eqnarray}\label{Hab}
H_{ab}&=&g^{cd}\sum_{\alpha,\beta=\pm}\sum_{l,m=1}^3\\
&&\left\{\left( M^{(\alpha,\beta,l,m)} G(X^{(\alpha,l)},X^{(\beta,m)})
+N^{(\alpha,\beta,l,m)} \epsilon(X^{(\alpha,l)},X^{(\beta,m)})
\right)X^{(\alpha,l)}_{c[a}X^{(\beta,m)}_{b]d}\right. \nonumber\\
&& + \left.\left(\widetilde{M}^{(\alpha,\beta,l,m)}G(X^{(\alpha,l)},\widetilde{X}^{(\beta,m)})+\widetilde{N}^{(\alpha,\beta,l,m)}\epsilon(X^{(\alpha,l)},\widetilde{X}^{(\beta,m)})\right)X^{(\alpha,l)}_{c[a}\widetilde{X}^{(\beta,m)}_{b]d} \right\},\nonumber
\end{eqnarray}
and
\begin{small}
\begin{eqnarray}
M^{(\alpha,\beta,l,m)}&=&\xi^{(\alpha,\beta,l,m)} |\lambda^{(\alpha,l)}|^{1/4}|\lambda^{(\beta,m)}|^{1/4} \left(\frac{|\lambda^{(\alpha,l)}|^{1/2}}{M_P}\right)^{c^{(\alpha,l)}}\left(\frac{|\lambda^{(\beta,m)}|^{1/2}}{M_P}\right)^{c^{(\beta,m)}},\\
N^{(\alpha,\beta,l,m)}&=&\chi^{(\alpha,\beta,l,m)} |\lambda^{(\alpha,l)}|^{1/4}|\lambda^{(\beta,m)}|^{1/4} \left(\frac{|\lambda^{(\alpha,l)}|^{1/2}}{M_P}\right)^{d^{(\alpha,l)}}\left(\frac{|\lambda^{(\beta,m)}|^{1/2}}{M_P}\right)^{d^{(\beta,m)}},\\
\widetilde{M}^{(\alpha,\beta,l,m)}&=&\widetilde{\xi}^{(\alpha,\beta,l,m)}|\lambda^{(\alpha,l)}|^{1/4}|\lambda^{(\beta,m)}|^{1/4}\left(\frac{|\lambda^{(\alpha,l)}|^{1/2}}{M_P}\right)^{\widetilde{c}^{(\alpha,l)}}\left(\frac{|\lambda^{(\beta,m)}|^{1/2}}{M_P}\right)^{\widetilde{c}^{(\beta,m)}},\\
\widetilde{N}^{(\alpha,\beta,l,m)}&=&\widetilde{\chi}^{(\alpha,\beta,l,m)}|\lambda^{(\alpha,l)}|^{1/4}|\lambda^{(\beta,m)}|^{1/4}\left(\frac{|\lambda^{(\alpha,l)}|^{1/2}}{M_P}\right)^{\widetilde{d}^{(\alpha,l)}}\left(\frac{|\lambda^{(\beta,m)}|^{1/2}}{M_P}\right)^{\widetilde{d}^{(\beta,m)}}.
\end{eqnarray}\end{small}
In these last expressions $\xi^{(\alpha,\beta,l,m)}$, $\widetilde{\xi}^{(\alpha,\beta,l,m)}$, $\chi^{(\alpha,\beta,l,m)}$, $\widetilde{\chi}^{(\alpha,\beta,l,m)}$, $c^{(\alpha,l)}$, $\widetilde{c}^{(\alpha,l)}$, $d^{(\alpha,l)}$ and $\widetilde{d}^{(\alpha,l)}$ are the free dimensionless parameters of the model with the restriction that
\begin{equation}
c^{(\alpha,l)},\widetilde{c}^{(\alpha,l)},d^{(\alpha,l)},\widetilde{d}^{(\alpha,l)}>-1/2,
\end{equation}
to get the correct limit when spacetime is flat. Note that equation (\ref{Hab}) does not include terms involving two eigenvectors with tilde. This is because these terms are equivalent, after a reparametrization of the free constants, to the terms that are present in equation (\ref{Hab}). Also observe that the power to which $M_P$ appears is a free parameter that has to be set by experiments.

This work shows that it is possible to investigate the phenomenological consequences of a Lorentz Invariant discrete spacetime structure at microscopic scales. The model is covariant, well-defined and, in principle, observable. In fact, it has been empirically tested in a high precision experiment \cite{Adelberger}, allowing to place bounds on some of its free parameters.

\section*{Acknowledgments}
This work was partially supported by the research grants CONACyT 101712
and PAPIIT-UNAM IN107412.

\end{document}